\documentclass{ws-ijmpe}
\usepackage[super,compress]{cite}
\begin{document}

\newcommand{\bef}{\begin{figure}}
\newcommand{\eef}{\end{figure}}
\newcommand{\nn}{\nonumber}
\newcommand{\be}{\begin{equation}}
\newcommand{\ee}{\end{equation}}
\newcommand{\bea}{\begin{eqnarray}}
\newcommand{\eea}{\end{eqnarray}}

\markboth{Subhash Singha, Bedangadas Mohanty and Zi-Wei Lin}{Studying
  re-scattering effect in heavy-ion collisions ...}

\catchline{}{}{}{}{}

\title{STUDYING RE-SCATTERING EFFECT IN HEAVY-ION COLLISION THROUGH $K^{*}$ PRODUCTION}

\author{Subhash Singha}

\address{Physics Department, Kent State University, Kent, OH 44242,USA\\
connectsubhash@gmail.com}

\author{Bedangadas Mohanty}

\address{School of Physical Sciences, National Institute of Science
  Education and Research, Jatni - 752050\\
bedanga@niser.ac.in}

\author{ Zi-Wei Lin}

\address{Department of Physics, East Carolina University, Greenville, NC
  27858-4353, USA\\
linz@ecu.edu}

\maketitle

\begin{history}
\received{Day Month Year}
\revised{Day Month Year}
\end{history}

\begin{abstract}
We have studied the $K^{*}$ production within A Multi-Phase Transport
model (AMPT) for Au+Au collisions at $\sqrt{s_{NN}}$ = 200 GeV to
understand the hadronic re-scattering effect on the measured yields
of the resonance. The hadronic re-scattering of the $K^{*}$ decay
daughter particles ($\pi$ and $K$) will alter their momentum distribution 
thereby making it difficult to reconstruct the $K^{*}$ signal
through the invariant mass method.  An increased hadronic re-scattering
effect thus leads to a decrease in the reconstructed yield of $K^{*}$
in heavy-ion collisions. Through this simulation study we argue that a
decrease in $K^{*}$/$K$ ratio with increase in collision centrality
necessarily reflects the hadronic re-scattering effect. Since the
re-scattering occurs in the hadronic phase and $K^{*}$ has a lifetime
of 4 fm/$c$, we present a toy model based discussion on using measured
$K^{*}$/$K$  to put a lower limit on the hadronic phase 
lifetime in high energy nuclear collisions. 
\end{abstract}


\ccode{PACS numbers:}


\section{Introduction}
High energy heavy-ion collisions form a system whose constituents
undertake various type of interactions during different times of
evolution of the system. Many resonances have been observed in these
collisions~\cite{Aggarwal:2010mt, Abelev:2008yz,Adams:2004ep,
  Abelev:2008zk, Abelev:2012hy} - $f_{2}(1270)$, $\rho(770)^{0}$,
$\Delta(1232)^{++}$,$f_{0}(980)$, $K^{*}(892)^{0\pm}$, $\Sigma(1385)$,
$\Lambda(1520)$ and $\phi(1020)$ with life times of 1.1 fm/$c$, 1.3 fm/$c$, 
1.6 fm/$c$, 2.6 fm/$c$, 4 fm/$c$, 5.5 fm/$c$, 12.6 fm/$c$ and 44
fm/$c$, respectively. Resonances are very good probes of the dynamics
of the system formed in heavy ion collisions~\cite{Brown:1991kk} as they cover from the
very early time scales to close to the freeze-out of the system. 

In the hadronic phase of the  system formed in heavy-ion collisions,
two important temperature or time scales comes into picture. One is 
the chemical freeze-out, where the inelastic collision among the 
constituents are expected to cease and the other is the later kinetic
freeze-out when the distance scales among the constituents are larger 
than the mean free path due to which all (elastic) interactions
cease~\cite{Rapp:2000gy, Song:1996ik, Rafelski:2001hp}.  If
resonances decay before kinetic freeze-out then they will be subject
to hadronic re-scattering of the daughter particles which will alter
their momentum distributions. This would lead to loss in the 
reconstruction of the parent resonance. The amount of loss could
depend on the life time of hadronic phase (specifically the time
between chemical and kinetic freeze-out), resonance daughter particle
hadronic interaction cross section, particle density in the medium and
the resonance phase space distributions. On the other hand after
chemical freeze-out pseudo-elastic interactions could regenerate
resonances in the medium leading to enhancement in their yields. For
example interactions like $\pi K \rightarrow K^{*0} \rightarrow \pi K$
could happen until kinetic freeze-out. Transport based model
calculations indeed predict that both re-scattering and re-generation
processes affect final resonance yields~\cite{Lin:2001cx}. On the other hand thermal
model calculations only after including re-scattering effects are able
to explain the experimentally measured ratios of resonance yield to
the yield of stable particles~\cite{Broniowski:2003ax, Rapp:2003ar}.

In this work we use A Multi Phase Transport Model (AMPT)~\cite{ampt,ampt2} to study the
effect of re-scattering on $K^{*}$ resonance production. The $K^{*}$
in this study denotes the sum of $K^{*\pm}$, $K^{*0}$ and $\bar{K^{*0}}$.
We have modified the AMPT code to track the production of $K^{*}$ 
(decayed and newly produced as a function of hadronic cascade time,
$\tau_{HC}$) and the change in the momentum distributions of its 
daughters ($\pi$ and $K$).  We demonstrate by gradually increasing
$\tau_{HC}$, that allows for  increased re-scattering effect, the invariant mass 
signal of $K^{*}$ reconstructed from the daughter particles ($\pi K$)
decreases. We present the change in the experimentally measured $K^{*}$ yield
per unit rapidity $dN/dy$ with the increase in $\tau_{HC}$.
We propose  an experimental observable the ratio of yield of $K^{*}$ to yield of
charged $K$ to understand the re-scattering effect in heavy-ion
collisions. Finally using the measured $K^{*}/K$ ratio, within the framework of
a simple model, we obtain the lower limit on the time difference
($\Delta t$)  between chemical and kinetic freeze-out in high energy 
heavy-ion collisions. 

The paper is organised as follows: in next section we briefly discuss
the AMPT model, in section III we discuss the results related to
re-scattering effects on the yield of $K^{*}$, it also includes a
discussion on the model dependent extraction of lower limit on time scale
between chemical and kinetic freeze-out and section IV summarises our
findings.

\section{AMPT Model}
The AMPT model~\cite{ampt,ampt2} uses the same initial conditions as in HIJING~\cite{hijing}.
However the minijet partons are made to undergo scattering before they are allowed
to fragment into hadrons. The AMPT model has two versions, one is the
default version, the other is called the string melting version.
All the results presented in this paper are based on the default
version (version 1.25t9b). 
The hadronic matter interaction is described by a hadronic cascade,
which is based on A Relativistic Transport (ART) model~\cite{Li:1995pr}. The termination time of the
hadronic cascade is varied in this paper from 0.6 fm/$c$ to 30 fm/$c$ to study the
effect of the hadronic re-scattering on the observables presented.
More detailed discussions regarding the AMPT model can be found in Ref.~\cite{ampt,ampt2}.
In this study, approximately 50000 events for each configuration
(different hadronic cascade time) were generated for Au+Au 0-80\% minimum bias collisions
at $\sqrt{s_{NN}}$ = 200 GeV. All results presented are for the
rapidity range $\mid y \mid$ $<$ 0.5.

$K^{*}$ resonances, together with their anti-particles, are included
explicitly~\cite{Lin:2001cx} in the hadronic cascade of the AMPT model. 
Both elastic and inelastic hadronic interactions involving $K^{*}$ are
included~\cite{ampt2} , such as $K^{*}$ productions, absorptions and decays. 
In particular, elastic scatterings of $K^{*}$ with a meson among ($\rho,\omega,\eta$)
are included using a 10 mb cross section, the same cross section as
used for kaon elastic scatterings~\cite{Li:1995pr}.
In addition to initial productions from the Lund string
fragmentation, the $K^{*}$ resonance can be produced from
kaon-pion scatterings, while $K^{*}$ decay is the inverse reaction. 
They can also be produced or destroyed from reactions such as 
$(\pi \eta) (\rho \omega) \leftrightarrow K^* \bar K$ or $\bar {K^*} K$, 
$\pi K \leftrightarrow K^* (\rho \omega)$~\cite{Brown:1991ig},
$\phi (\pi \rho \omega) \leftrightarrow (K K^*)(\bar K \bar {K^*})$, and
$\phi (K K^*) \leftrightarrow (\pi \rho \omega) (K K^*)$~\cite{Alvarez-Ruso:2002ib}.


\section{Results}

\subsection{Invariant mass distribution}

\bef
\begin{center}
\includegraphics[scale=0.4]{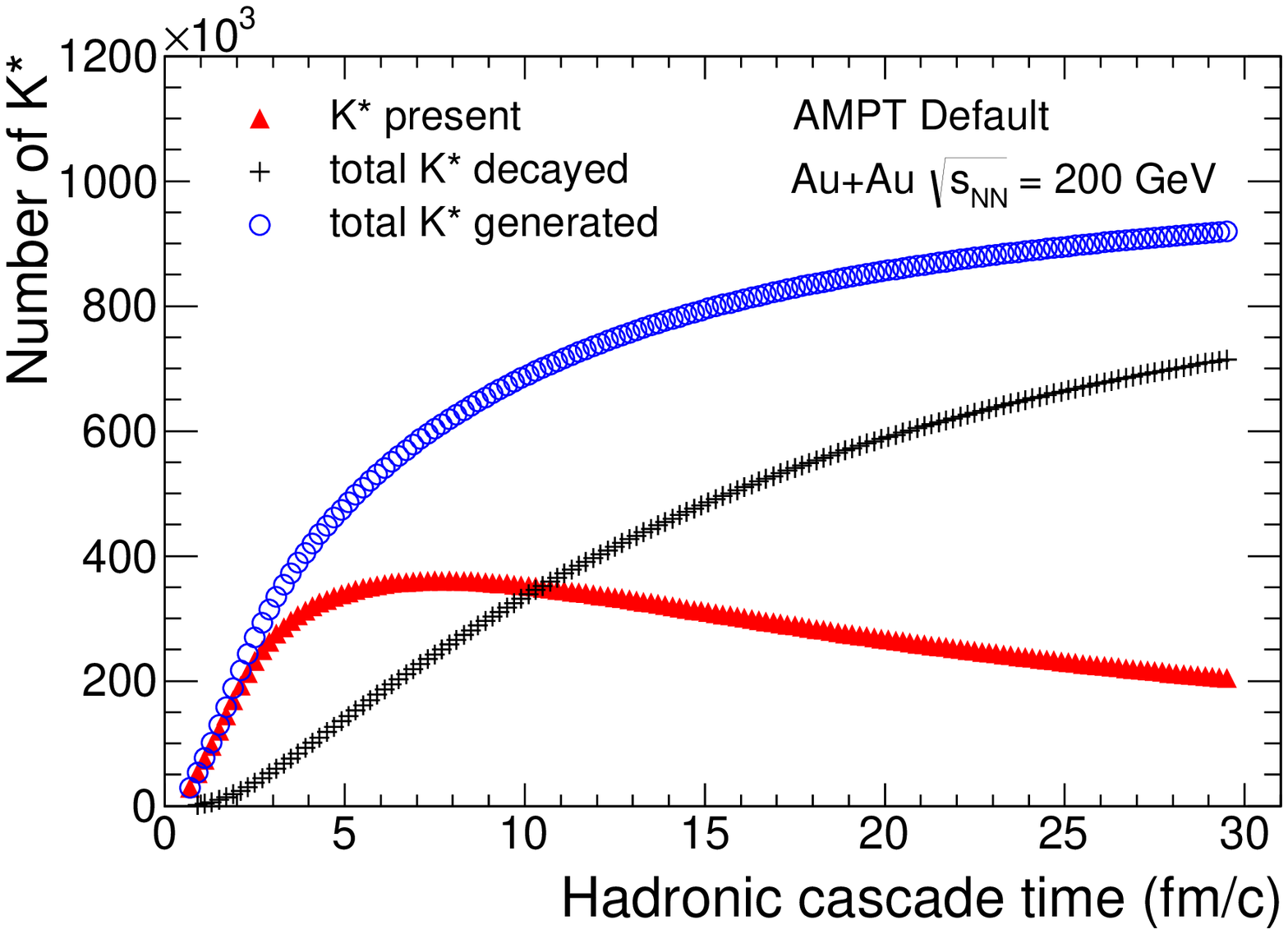}
\caption{(Color online) Number of $K^{*}$ as a function of hadron
  cascade time in the default version of the AMPT model for Au+Au collisions at $\sqrt{s_{NN}}$ =
  200 GeV. Red solid triangle corresponds to total $K^{*}$ present
at any given hadron cascade time. Blue open circle corresponds to total  $K^{*}$
produced. Black cross corresponds to total $K^{*}$ decayed. }
\label{fig1}
\end{center}
\eef
Figure~\ref{fig1} shows the time evolution of $K^*$ yields for minimum bias
Au+Au at collisions $\sqrt {s_{NN}}$=200GeV from the default version
of AMPT.  
Red solid triangles represent the total number of $K^*$ present at a
given time in the hadron cascade. Open blue circles
represent the total number of produced $K^*$, 
black crosses represent the total number of decayed $K^*$, 
while their difference corresponds to the total number of $K^*$
present at the time. 
We see that the number of $K^*$ present reaches a peak after several
$fm/c$ (partly due to the finite formation time of hadrons) and then slow
decreases with time; it will eventually vanish at large enough time as the
system expands and resonances decay away. 

The $\pi K$ daughters of the $K^{*}$ decayed
will undertake re-scattering effects which is expected to increase with increase in $\tau_{HC}$.
The hadronic re-scattering of daughter $\pi K$ would then lead to loss in
$K^{*}$ invariant mass signal . The
regeneration of  $K^{*}$  though will also pick up with increase in hadron
cascade time, however if not dominant will not be sufficient to compensate the loss due to
re-scattering. The $K^{*}$ regeneration depends on the the cross
section $\sigma_{K\pi}$
while the re-scattering of daughter particles depends on cross sections
$\sigma_{\pi\pi}$ and $\sigma_{K\pi}$, where $\sigma_{\pi\pi}$ are considerably larger
(factor $\sim$ 5) than $\sigma_{K\pi}$~\cite{Protopopescu:1973sh, Matison:1974sm, Bleicher:1999xi}.

\bef
\begin{center}
\includegraphics[scale=0.4]{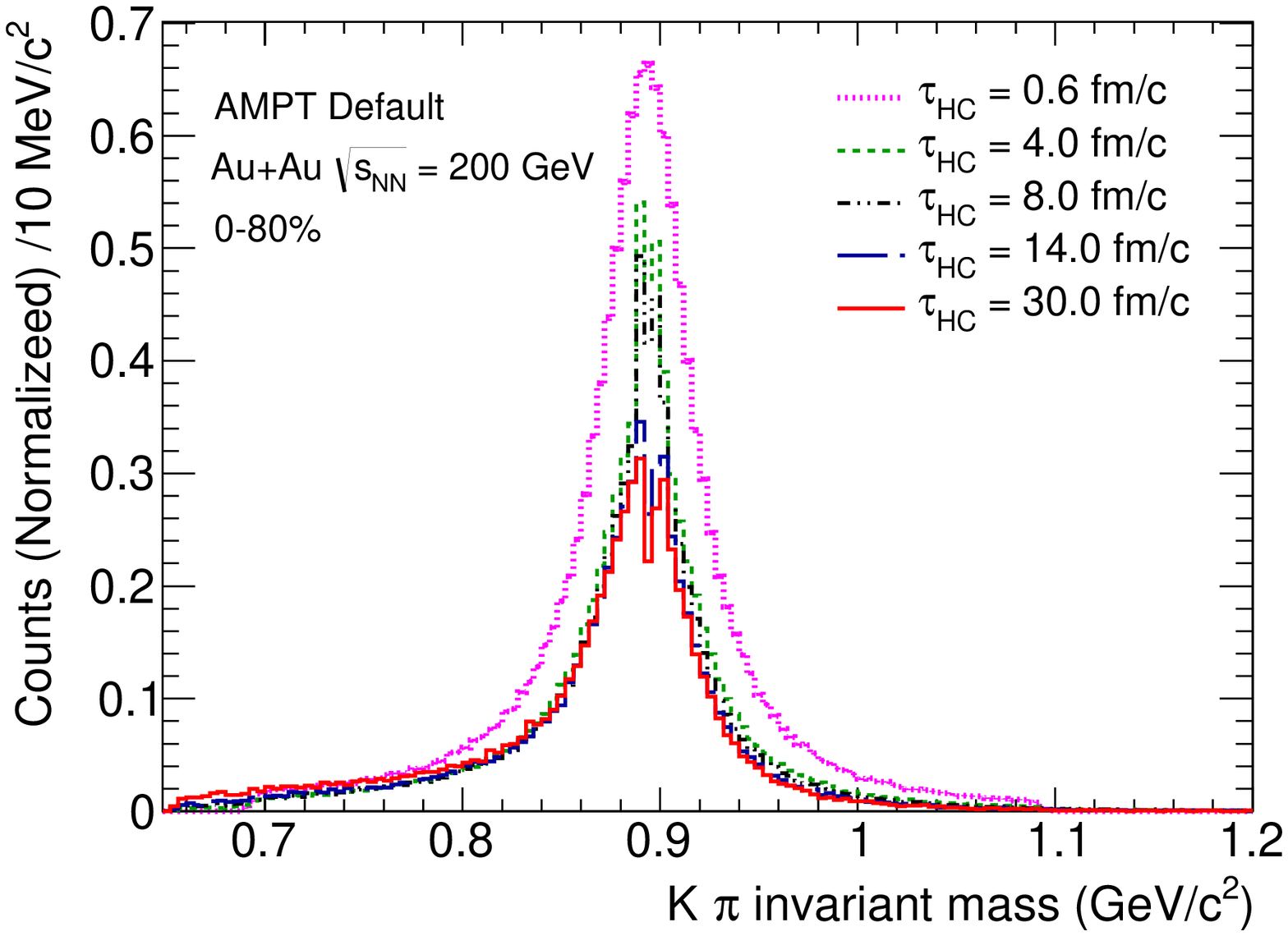}
\caption{(Color online) $K^{*}$  $\rightarrow$ $\pi K$ invariant
  mass distribution in Au+Au collisions at $\sqrt{s_{NN}}$ = 200 GeV
  from the AMPT model.
The different distributions correspond to different termination time
of hadron cascade ($\tau_{HC}$). The number of events are kept the
same for all the
configurations.}
\label{fig2}
\end{center}
\eef
Figure~\ref{fig2} shows the invariant mass of $K^{*}$ meson
reconstructed using four momentum information of the $\pi$ and $K$
in minimum bias  (0-80\% of the total cross section) Au+Au collisions at $\sqrt{s_{NN}}$ = 200 GeV. The
results are shown for different configurations of AMPT calculations
where the termination time of the hadron cascade is varied from 0.6 to
30 fm/$c$. The number of events generated for each configurations are kept
the same in order to make proper comparisons. Clearly one observes a
decrease in the invariant mass signal as the hadron cascade time
increases. An increase in hadron cascade time corresponds to increase
in hadronic re-scattering effects in the AMPT model. The loss in $K^{*}$
signal strength is anticipated due to the change in momentum of the 
daughter $\pi$ and $K$ of  $K^{*}$ meson as a result of hadronic
re-scattering. However we have observed that the reconstructed $K^{*}$ mass and
width are not affected by the hadronic re-scattering process as
implemented in the model.

\subsection{Observables for re-scattering effect}
\bef
\begin{center}
\includegraphics[scale=0.4]{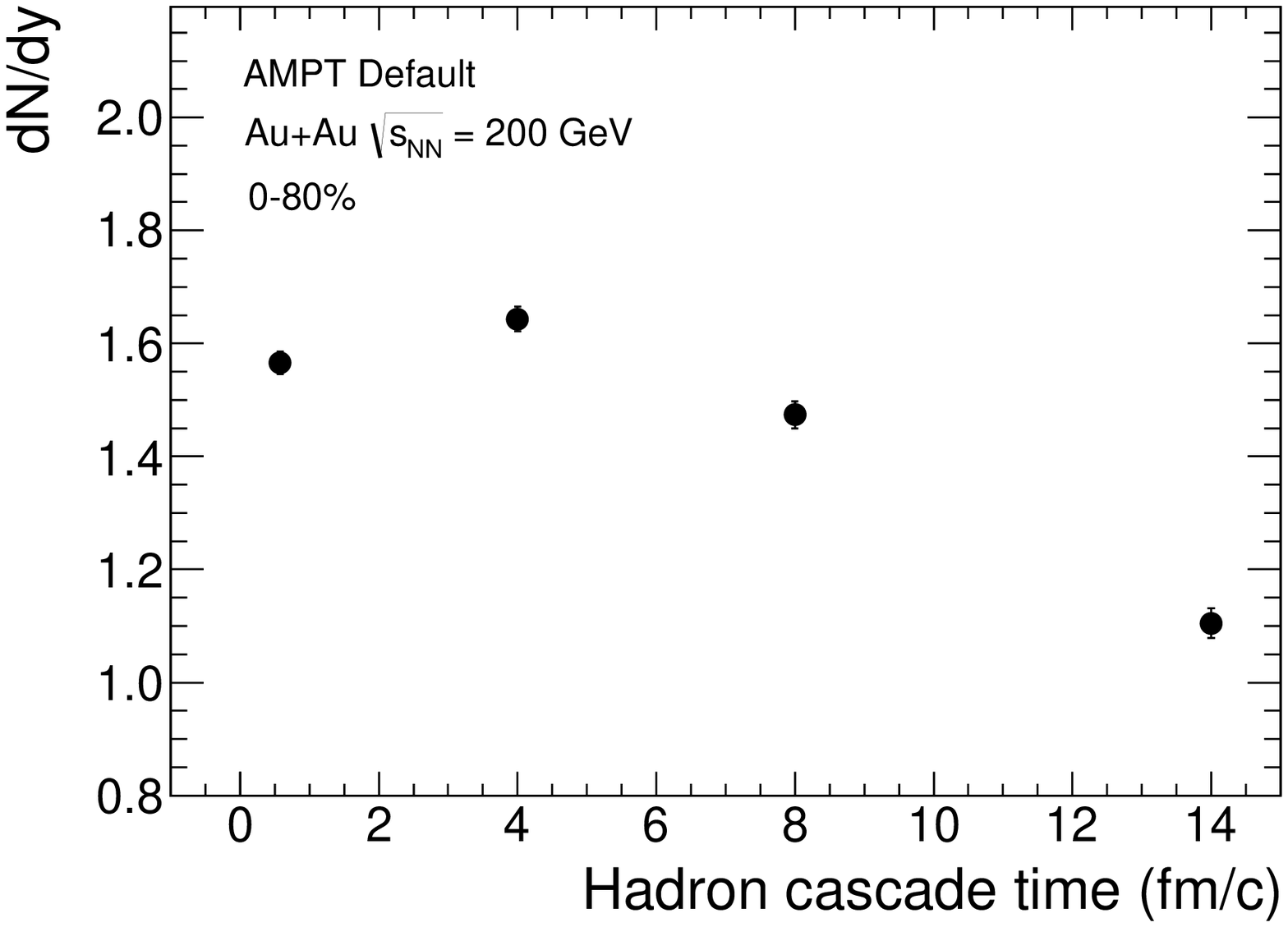}
\includegraphics[scale=0.4]{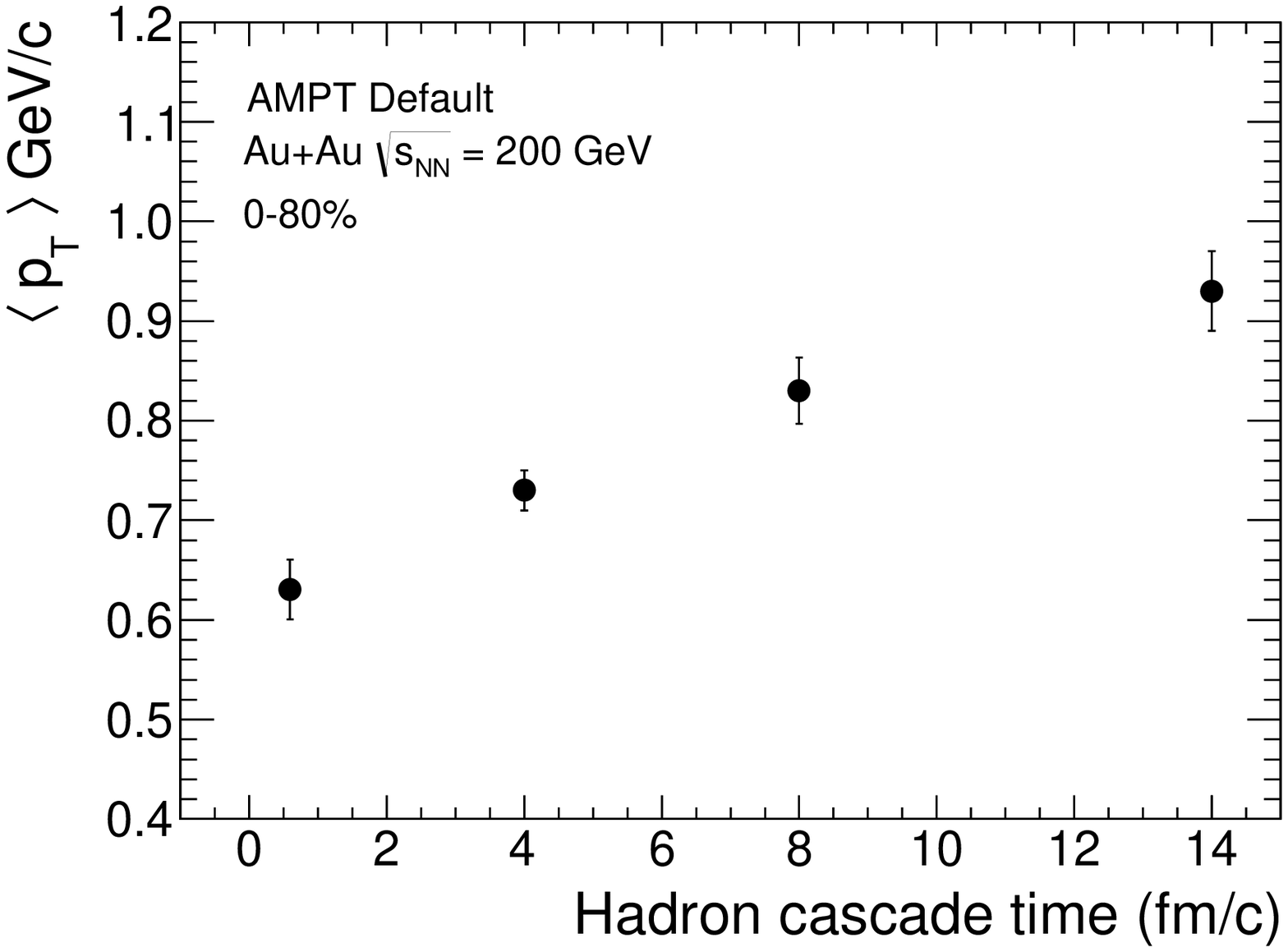}
\caption{ $dN/dy$ (upper panel) and $\langle p_{T} \rangle$)  (lower
  panel) of $K^{*}$ meson in AMPT model for Au+Au collisions at
  $\sqrt{s_{NN}}$ = 200 GeV as a function of hadron cascade time.}
\label{fig3}
\end{center}
\eef
In this sub-section we discuss how to quantify the re-scattering
effect in heavy-ion collisions. Figure~\ref{fig3} shows the $dN/dy$
(upper panel) and mean transverse momentum ($\langle p_{T} \rangle$)
of the reconstructed $K^{*}$ as a function of hadron cascade time. As
expected the $dN/dy$ decreases due to re-scattering effects as hadron
cascade time increases. We find the $\langle p_{T} \rangle$
values to increase as a function of hadron cascade time, this is also
consistent with the loss of $K^{*}$ due to re-scattering. This could
also mark the increase of transverse flow with time. 

\bef
\begin{center}
\includegraphics[scale=0.4]{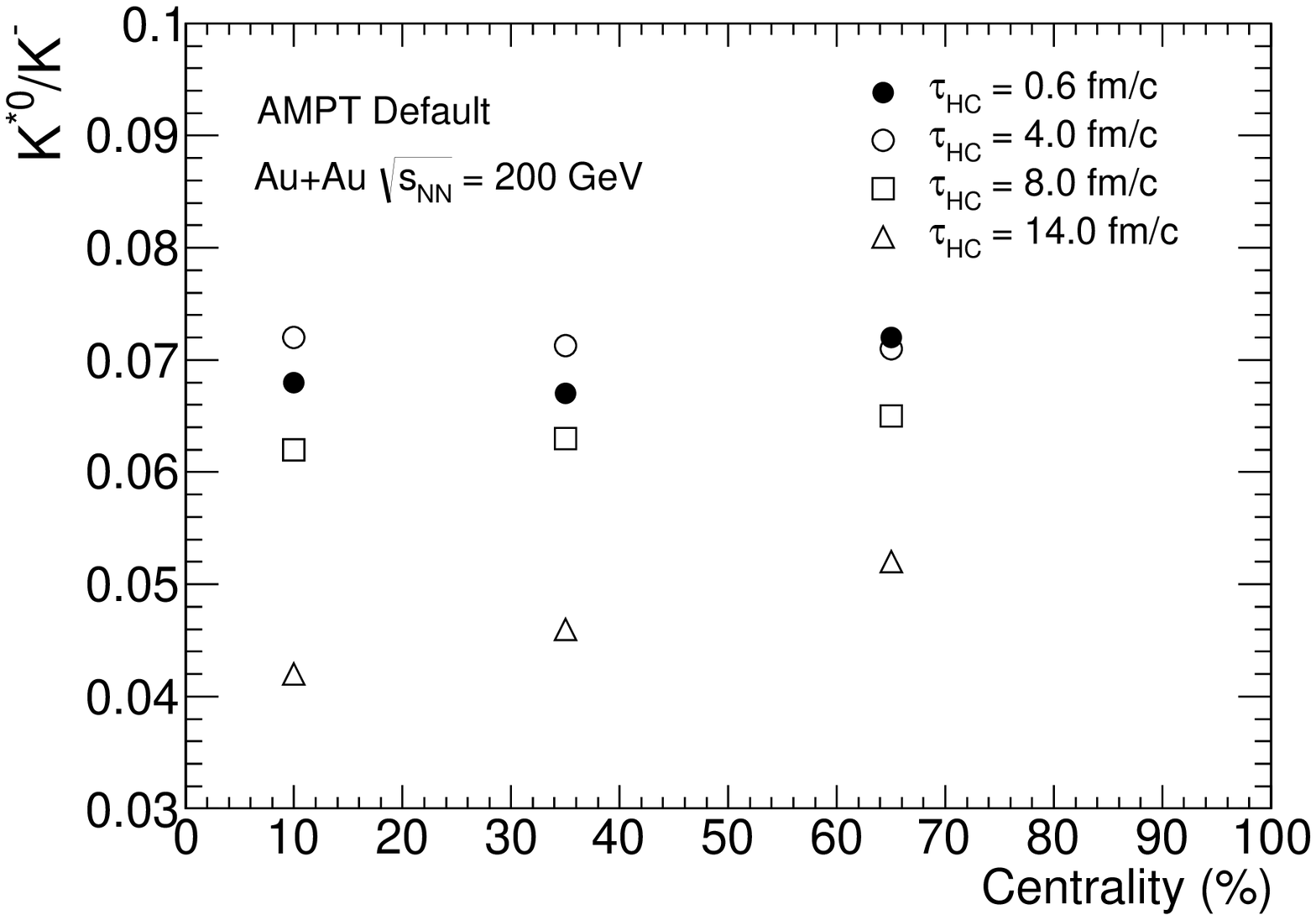}
\includegraphics[scale=0.4]{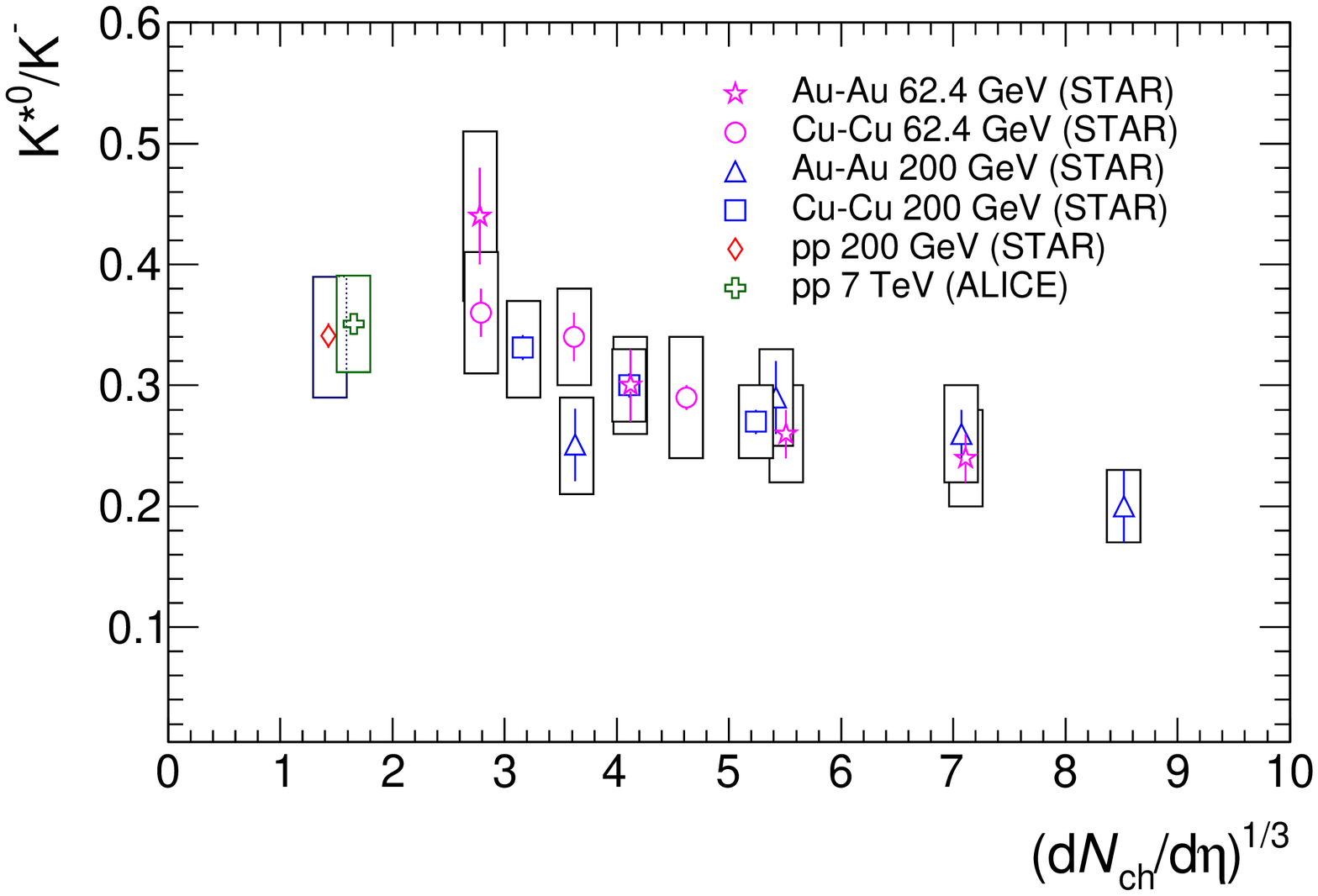}
\caption{(Color online) Upper panel: $K^{*0}/K^{-}$ versus
  collision centrality in AMPT model for Au+Au collisions at
  $\sqrt{s_{NN}}$ = 200 GeV. Results are shown for different hadron
  cascade time. Lower panel: $K^{*0}/K^{-}$ as a function of
  $(dN_{ch}/d\eta)^{1/3}$ from experimental data in heavy-ion and p+p collisions~\cite{Aggarwal:2010mt, Abelev:2008yz,Adams:2004ep,
  Abelev:2008zk, Abelev:2012hy}. The boxes represent systematic errors
and vertical lines represent the statistical errors. }
\label{fig4}
\end{center}
\eef
However in order to provide a proper observable in an experiment for re-scattering we
need an appropriate baseline to compare. One such observable could be the
ratio of yield of $K^{*}$ to $K^{-}$
($K^{*}/K^{-}$). Figure~\ref{fig4}  (upper panel) shows the
$K^{*0}/K^{-}$ versus collision centrality for different hadron
cascade time. One observes a clear decrease in the value of the ratio 
with increase in the hadron cascade time. Indicating that the ratio
is sensitive to re-scattering effect. At large hadron cascade time,
the ratio shows a decrease from peripheral collisions to central
collisions. The central collisions are expected to live longer and
provide a medium with higher possibility of re-scattering. Hence a
clear centrality
dependence of $K^{*}/K^{-}$ or a decrease in the value of $K^{*}/K^{-}$
with respect to $p$+$p$ collisions would indicate dominance of
re-scattering effect in heavy-ion collisions. Regeneration on the
other hand would lead to increase of the ratio from peripheral to
central collisions and with respect to $p$+$p$
collisions. Figure~\ref{fig4} (lower panel) shows the
compilation of $K^{*0}/K^{-}$ versus $(dN_{ch}/d\eta)^{1/3}$
experimental data from heavy-ion and $p$+$p$ collisions~\cite{Aggarwal:2010mt, Abelev:2008yz,Adams:2004ep,
  Abelev:2008zk, Abelev:2012hy}. One observes that
$K^{*0}/K^{-}$  in p+p collisions is larger than in central
heavy-ion collisions. There is a clear centrality (reflected by the
values of $(dN_{ch}/d\eta)^{1/3}$, where $dN_{ch}/d\eta$ is the number of
charged particles per unit pseudo-rapidity) dependence of the ratio, thereby
indicating presence of hadronic re-scattering in heavy-ion
collisions.

\section{Hadronic phase time}

\bef
\begin{center}
\includegraphics[scale=0.4]{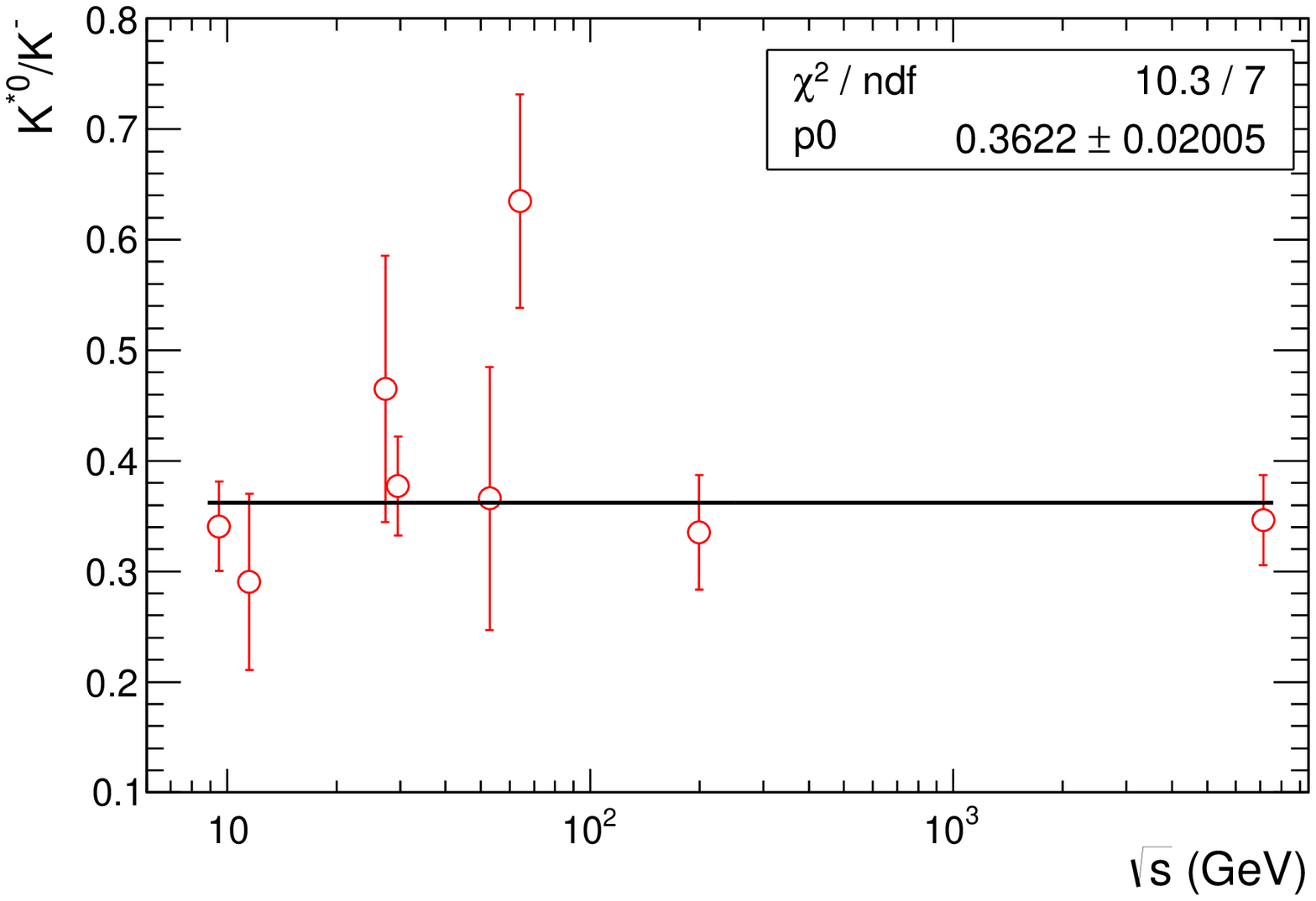}
\includegraphics[scale=0.4]{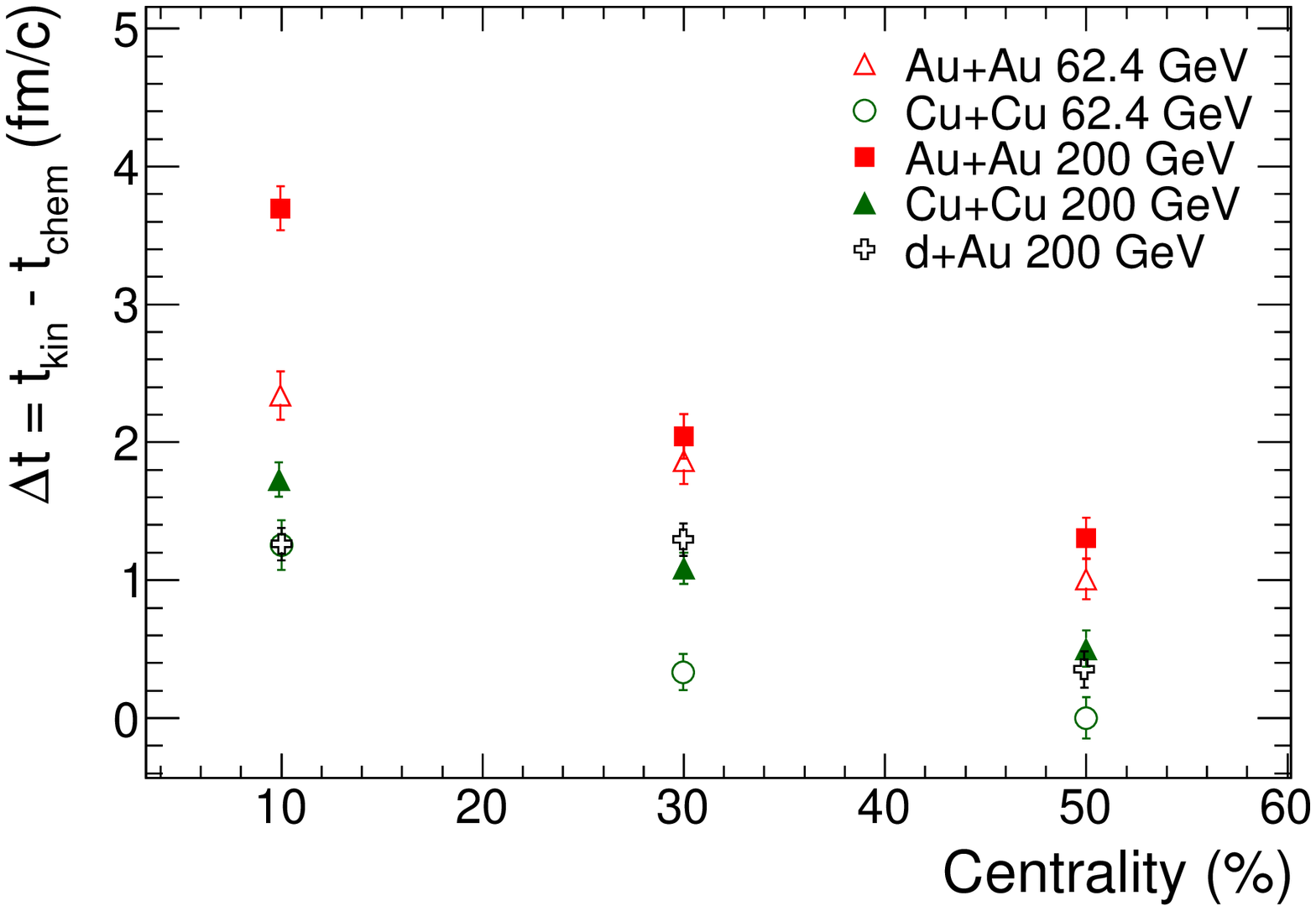}
\caption{(Color online) Upper panel: $K^{*0}/K^{-}$  for $p$+$p$
  collisions at various center of mass energies. Lower panel: Estimate
  of the time difference between chemical and kinetic freeze-out for
  various collision systems and beam energies as a function of
  collision centrality. See text for details.}
\label{fig5}
\end{center}
\eef
The suppression in $K^{*}/K^{-}$  ratio in heavy-ion collisions
with respect to $p$+$p$ collisions can be used to set a lower limit on the
time difference between chemical and kinetic freeze-out.  The
experimental results shows that $K^{*}/K^{-}$ decreases with 
increase in collision centrality. This implies that $K^{*}$
re-scattering is dominant over $K^{*}$ regeneration. This in turn means
that $K^{*} \leftrightarrow K\pi$ is not in balance. Hence one can in
principle use the $K^{*}/K^{-}$  to get a lower limit estimate
of the time between chemical and kinetic freeze-out as,
$[K^{*}/K^{-}]_{kinetic}$ = $[K^{*}/K^{-}]_{chemical}$
$\times$ $e^{-\Delta t/\tau}$. Where $\tau$ is the $K^{*}$ life time,
taken here as 4 fm/$c$ ignoring any medium modification of the
width of invariant mass distribution of $K^{*}$, supported by the
experimental data~\cite{Aggarwal:2010mt, Abelev:2008yz,Adams:2004ep,
  Abelev:2008zk, Abelev:2012hy}.  $\Delta t$ is the time between the chemical and
kinetic freeze-outs. We assume that the
$[K^{*}/K^{-}]_{chemical}$ is given by the experimental data in
$p$+$p$ collisions and the heavy-ion collision data provides the
$[K^{*}/K^{-}]_{kinetic}$. This is equivalent to assuming that 
all $K^{*}$ which decay before kinetic freeze-out are lost
due to re-scattering effect and there is no regeneration effect between
kinetic and chemical freeze-out. AMPT model simulations (Fig.~\ref{fig2}) shows this
assumption could hold to a substantial
extent. However these assumptions can only make the estimates of
$\Delta t$ to be 
a lower limit for the the time difference between chemical and kinetic
freeze-outs.  Figure~\ref{fig5} upper panel shows the value of
$[K^{*}/K^{-}]_{chemical}$ taken from the available experimental p+p data to
be 0.36 and the heavy-ion data (corresponding to
$[K^{*}/K^{-})_{kinetic}$ ) is taken from Fig.~\ref{fig4}. The
results for $\Delta t$ boosted by the Lorentz factor ($\sim$ 1.38-1.57) for three different centralities for various
systems are plotted in the lower panel of Fig.~\ref{fig5}. We find  the time difference between chemical and kinetic freeze-out
increases with both beam energy and system size as expected. 
For the central most Au+Au collisions at $\sqrt{s_{NN}}$ = 200 GeV, the lower limit of time between chemical
and kinetic freeze-out is about 3.7 fm/$c$.

\section{Summary}
 In summary, we have provided a detail study of effect of hadronic
 re-scattering on $K^{*}$ production using the AMPT model. The study
 has been carried out by varying the termination time of the hadronic
 cascade. Larger the hadronic cascade time more is the
 re-scattering of the daughters ($\pi K$) of the $K^{*}$ meson. We observe
 that the reconstructed $K^{*}$ signal is lost due to re-scattering of
 the daughters which results in the change in their momentum
 distributions. There is a clear decrease in $dN/dy$ of the
 reconstructed $K^{*}$ meson with increase in hadron cascade time and 
 the $\langle p_{T} \rangle$ increases with hadron cascade time. We propose
 an observable $K^{*}/K^{-}$ as a
 function of collision centrality to study the re-scattering effect in
 heavy-ion collisions. A clear decrease in the $K^{*}/K^{-}$
 ratio with respect to p+p collisions and with increase in collision
 centrality can be considered as a  signature of re-scattering
 effect in heavy-ion collisions. Within the framework of a toy model,
 it was possible to use the measured $K^{*}/K^{-}$ ratio in $p$+$p$
 and A+A collisions to estimate the lower limit of the time difference between chemical
 and kinetic freeze-out. For the most central collisions
 at RHIC this lower limit of the time difference is found to be about
 3.7 fm/$c$, a value which is consistent with other estimates~\cite{Adams:2003xp,Adams:2003ra}.

\section*{Acknowledgements}

BM is supported by the DAE-BRNS project sanction
No. 2010/21/15-BRNS/2026. \\

\end{document}